# A Framework for SLO, Carbon, and Wastewater-Aware Sustainable FaaS Cloud Platform Management

Sirui Qi, Hayden Moore
*Colorado State University*
Fort Collins, USA
{alex.qi, hayden.moore}@colostate.edu

Ninad Hogade, Dejan Milojicic, Cullen Bash
*Hewlett Packard Labs*
Milpitas, USA
{ninad.hogade, dejan.milojicic, cullen.bash}@hpe.com

Sudeep Pasricha
*Colorado State University*
Fort Collins, USA
sudeep@colostate.edu

*Abstract—* Function-as-a-Service (FaaS) is a growing cloud computing paradigm that is expected to reduce the user cost of service over traditional serverful approaches. However, the environmental impact of FaaS has not received much attention. We investigate FaaS scheduling and scaling from a sustainability perspective in this work. We find that the service-level objectives (SLOs) of FaaS and carbon emissions conflict with each other. We also find that SLO-focused FaaS scheduling can exacerbate water use in a datacenter. We propose a novel sustainability-focused FaaS scheduling and scaling framework to co-optimize SLO performance, carbon emissions, and wastewater generation.

## I. Introduction

Function-as-a-Service (FaaS) is a fast growing model of cloud computing with many benefits over traditional serverful approaches. In FaaS, developers create applications as a set of functions that are packaged in containers for deployment in cloud platforms. FaaS relieves developers from the burden of managing servers, scaling, logging and fault handling, which are now handled by the cloud provider. Several commercial FaaS services have emerged, including AWS Lambda and Google Cloud Functions. The fine granularity of function execution enables lower costs for developers who only pay when a function is executed. But unlike virtual machines (VMs) in serverful approaches, cloud providers now need to more carefully schedule many more finer granularity containers to execute functions while meeting their service level objectives (SLOs) and scale up containers to meet request spikes. However, current estimates indicate that in cloud datacenters, about 50% of energy is used by idle resources, highlighting the inadequate management of these platforms [1].

Beyond improving SLO performance guarantees, there is also concern about the environmental impact of cloud datacenters. Currently datacenters consume approximately 1–2% of global electricity demand [2]. By 2030, this is projected to increase to between 3–13% [3]. Such a massive increase in electricity usage will proportionally increase the carbon footprint of these datacenters, as fossil fuels are still widely used for energy production. Moreover, cloud datacenters also consume significant water for cooling and as part of their energy generation footprint. Currently datacenters consume 626 billion liters of water a year in the United States [4] and water scarcity is becoming an increasing concern globally (about 30% of cities globally with a population above a million experience yearly water scarcity [5]). As FaaS grows in popularity, there is an urgent need to focus on making FaaS cloud datacenter platforms more carbon and water efficient.

Several efforts in recent years have focused on improving the energy efficiency of FaaS. Much of the focus has been on reducing the energy usage of FaaS cloud platforms [6]. But energy usage does not directly correlate to carbon or water emissions [2]. A recent approach proposed a carbon-aware FaaS scheduler [1], which exploits geo-spatial differences in carbon intensity to migrate FaaS requests to greener locations. However, significant SLO violations can occur due to the high latency associated with moving functions between datacenters.

In this paper, we present the first framework that simultaneously co-optimizes function SLOs, operational carbon, and wastewater generation in FaaS cloud datacenter environments. We propose a novel multi-objective optimization framework and demonstrate improvements over the state-of-the-art FaaS scheduling frameworks.

## II. Sustainable FaaS Management Framework Overview

We focus on the problem of making scheduling (mapping containers to compute nodes) and autoscaling (provisioning compute cores and memory to each container) decisions for containers that execute incoming functions at a FaaS cloud datacenter facility. The decisions are made at the start of a developer defined epoch, based on the currently pending functions and predictions of functions projected to arrive during the epoch. The functions can have different execution times and diverse (SLO-based) deadlines. The goal of our framework is to co-optimize function execution in this FaaS cloud, minimizing SLO violations, carbon emissions, and wastewater generation at the same time.

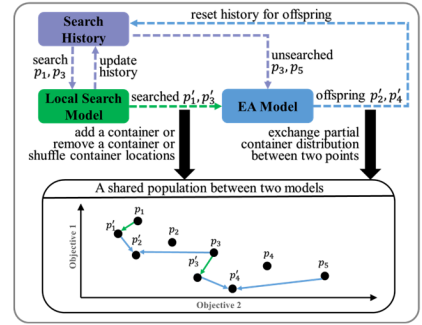

**Fig. 1 Sustainable FaaS cloud management framework overview**

Fig. 1 shows an overview of the algorithmic approach used in our proposed sustainable FaaS cloud management (*SFCM*). *SFCM* uses a randomized population for starting points. Each starting point $p$ represents a unique container scheduling and autoscaling plan in the population. The scheduling plan determines the number of containers per function ID (type) and the number of requests allocated to each container, while the autoscaling plan determines the size of each container (number of cores, memory). It is possible to map multiple containers to the same server node to improve resource utilization.

The framework consists of three main components: search history, local search heuristic model, and a global evolutionary algorithm (EA) model. Fig. 1 illustrates a simplified example with 5 points in the population (i.e. 5 valid solutions) and 2 objectives to optimize. The local search model uses a heuristic that attempts to 1) add containers, 2) remove containers, or 3) shuffle current container locations of a random function ID to find neighboring points in the search (solution) space. If the neighboring point is better (in terms of a weighted sum of the metrics) than the starting point, it will replace the starting point and continue with local searching. After local search, the searched points ($p'_1, p'_3$ in the figure) are passed to the EA model to combine with unsearched point $p_3, p_5$ to generate offsprings $p'_2, p'_4$, which can replace any dominated points $p_2, p_4$ in the population. The EA model uses evolutionary search techniques to evolve the population of solutions, with help from the local search model. The search history component (see Fig. 1) records the update frequency of each point (solution in the search space) considered by the local search model. It helps the local search model pick search starting points with the highest update frequency. Meanwhile, the EA model resets the search history back to the default value of corresponding starting points if replaced by offsprings. In this manner, the search history boosts local search performance during design space exploration and works with EA to jump out of local optima. The co-operation between local search and EA helps explore the complex design space of FaaS scheduling and scaling to meet the dynamic needs of function execution in FaaS cloud contexts.

## III. FaaS Workload and Modeling Assumptions

To evaluate our framework, we consider a two-week serverless function production trace from Microsoft's Azure service [7]. Based on their sanitized original traces, we derived our serverless function traces with 424 unique serverless function IDs, each of which represents a unique execution functionality and runtime. A container can process multiple requests from the same function ID in parallel but may require larger CPU resource allocations. Meanwhile, when a container starts up, shuts down, or stays idle, it always introduces resource usage and delays in the cluster, which are modeled as configurable variables in the experiment, derived from a real FaaS cluster system. The number of function IDs in each epoch varies from 13 to 62, with a runtime distribution where nearly 90% of functions execute for less than 30 seconds. Each function has an SLO-based deadline, which if missed, results in an SLO violation.

We carefully model the carbon emissions and wastewater generation based on the models in [8], for our FaaS cloud datacenter cluster. The carbon emissions of a cluster are based on the total energy consumption and water usage. This is because during the processes of 1) electricity generation, 2) potable water production, and 3) wastewater treatment, "dirty" energy sources are utilized and carbon dioxide is released into the atmosphere. The carbon emissions of a cluster are also impacted by real-time factors such as cooling efficiency, carbon intensity of electricity, etc. Meanwhile, a large amount of water is consumed in the cooling units of datacenters to transfer the heat from a compute room to open space through water cycling and evaporation. Cycling water needs to be replaced regularly and evaporated water is directly released into the air. If there are thermal hotspots in the datacenter, more cooling energy is expended. We consider all of these effects when modeling carbon emissions and wastewater generation in our FaaS cloud cluster.

## IV. Experimental Results

We consider a 50 node FaaS cluster, with each node consisting of 128 cores (2 EPYC 7713 CPUs with 64-cores each) and 256 GB RAM. We select two state-of-the-art FaaS management frameworks to compare with our proposed (*SFCM*) framework. The first work (*SCORE*) utilizes a scoring system to schedule and scale containers and improves on the default Kubernetes scheduling algorithm [9]. The second work (*HYBRID*) proposes a hybrid method which utilizes both non-stopping virtual machines for long-running functions and short-lived containers for short-running functions [10]. All three frameworks are adapted to make decisions every 15 minutes (epoch length) for different request intensity and functions IDs from the Azure function trace, discussed in the previous section. Both the container locations and sizing (scaling) decisions are determined by frameworks for each function ID in each epoch.

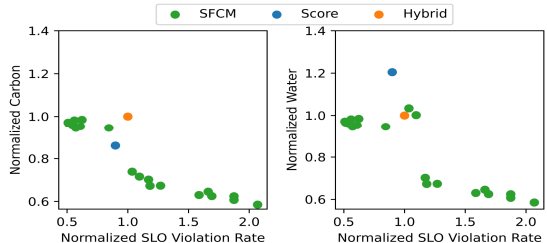

**Fig. 2. Pareto front of our multi-objective framework vs. [9], [10]**

First, we illustrate the generated solutions from the two state-of-the-art FaaS scheduling frameworks *SCORE* [9] and *HYBRID* [10] along with solutions generated by our multi-objective framework *SFCM* in one epoch, in Fig. 2 (with the *HYBRID* approach's objective values as normalization baseline). Instead of showing the 3-objective Pareto front (carbon, water, SLO) which is difficult to visualize in 3D, we show two simpler plots: a 2D Pareto front of carbon emission and SLO violations on the left, and a 2D Pareto front of water usage and SLO violation rate on the right. From Fig. 2, we can clearly see that *SFCM* provides a diverse solution set unlike *SCORE* and *HYBRID*.

Furthermore, most solutions from *SFCM* dominate solutions from the two state-of-the-art frameworks.

Lastly, Fig. 3 shows results aggregated over an 8-hour interval. We show results for four variants of our *SFCM* framework, where we pick solutions that optimize SLO (*SFCM-SLO*), carbon (*SFCM-Carbon*), water (*SCFM-Water*), and a weighted sum of the 3 metrics (*SFCM-Balance*). It can be observed that *SFCM-SLO*, *SFCM-Carbon*, and *SFCM-Water* provide up to 22%, 35%, and 37% reduction in corresponding objectives compared to the best state-of-the-art framework (*SCORE*). *SFCM-Balance* attempts to co-optimize across all three objectives. It dominates the *HYBRID* framework, with 45%, 25%, and 26% lower SLO violation rate, carbon emissions, and water usage. Compared with *SCORE*, *SFCM-Balance* is able to reduce carbon and water use by 14% and 20% respectively, while only compromising <1% SLO violation rate.

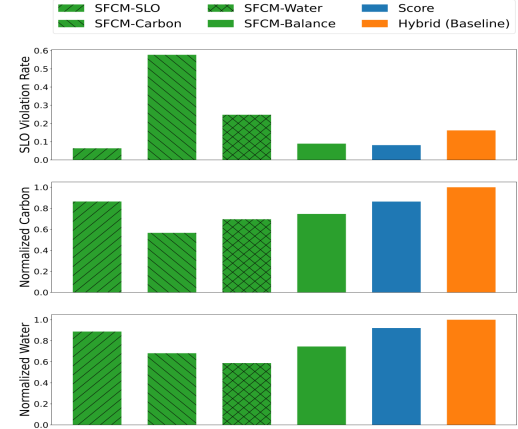

**Fig. 3. SLO violation rate (%), carbon usage, and wastewater generation results for the proposed SFCM framework vs [9], [10]**

## V. Conclusion

In this paper, we studied the problem of sustainable FaaS container scheduling and autoscaling in a single cloud datacenter to simultaneously minimize SLO violation rate, carbon emissions, and wastewater generation. We developed a novel multi-objective optimization framework (*SFCM*) for this problem. Our experimental results show that *SFCM* provides a diverse set of solutions, many of which dominate the solutions produced by state-of-the-art FaaS scheduling frameworks. In an 8-hour cumulative test on a FaaS cloud cluster, our framework reduces SLO violation rate, carbon emissions, and water usage by up to 45%, 25%, and 26% correspondingly compared to state-of-the-art FaaS scheduling frameworks.